\documentclass[12pt,a4]{article}
\usepackage{sw20lart}

\input tcilatex
\begin{document}

\title{\textbf{Algebraic Characterization of Vector Supersymmetry in Topological
Field Theories }}
\author{\textbf{L.C.Q.Vilar, O. S. Ventura, C.A.G. Sasaki } \\
CBPF, Centro Brasileiro de Pesquisas F\'{\i}sicas \\
Rua Xavier Sigaud 150, 22290-180 Urca \\
Rio de Janeiro, Brazil\vspace{2mm}\\
and\vspace{2mm} \and \textbf{S.P. Sorella} \\
UERJ, Universidade do Estado do Rio de Janeiro\\
Departamento de F\'\i sica Te\'orica\\
Instituto de F\'{\i}sica\\
Rua S\~ao Francisco Xavier, 524\\
20550-013, Maracan\~{a}, Rio de Janeiro, Brazil \and \textbf{CBPF-NF-007/97}%
\vspace{2mm}\newline \and \textbf{PACS: 11.10.Gh}}
\maketitle

\begin{abstract}
An algebraic cohomological characterization of a class of linearly broken
Ward identities is provided. The examples of the topological vector
supersymmetry and of the Landau ghost equation are discussed in detail. The
existence of such a linearly broken Ward identities turns out to be related
to BRST exact antifield dependent cocycles with negative ghost number.

\setcounter{page}{0}\thispagestyle{empty}
\end{abstract}

\vfill\newpage\ \makeatother
\renewcommand{\theequation}{\thesection.\arabic{equation}}

\section{\ Introduction\-}

The topological theories \cite{bbrt} are known to be characterized, besides
their BRST simmetry, by the so called \textit{topological vector\
supersymmetry }\cite{brt,dgs,gms,br}; an additional invariance possessing
rather interesting properties.

The first property is that the generators of the topological susy carry a
Lorentz index and, together with the BRST operator, give rise to an algebra
of the Wess-Zumino type which, closing on-shell on the space-time
translations, allows for a supersymmetric interpretation \cite
{brt,dgs,gms,br}.

The second feature of the topological susy is that it is present only after
the introduction off all the ghost fields needed in order to quantize the
model, \textit{i.e. }it is an invariance of the fully quantized action and
not only of its classical part as one can check, for instance, in the case
of the Schwartz type topological $BF$ models \cite{gms}. It should also be
remarked that this last feature is unavoidably related to the choice of the
gauge fixing term. In other words, the vector susy can exist only for
certain values of the gauge parameters present in the gauge fixing
condition. As an example of this feature let us mention the three
dimensional Chern-Simons model for which, among the class of the linear
covariant gauge fixings, the vector susy turns out to select the Landau
gauge \cite{brt,dgs}.

The third interesting aspect of the vector susy is that, after the
introduction of the antifields (or BRST external sources), the algebra
between the vector susy Ward operator and the BRST operator closes off-shell
on the space-time translations, without making use of the equations of
motion. Moreover, the vector susy loses now the property of being an exact
invariance of the fully quantized action. Rather, it yields a broken Ward
identity \cite{dgs,gms,br}. It is a remarkable feature, however, that the
corresponding breaking term is in fact a classical breaking, \textit{i.e. }a
breaking which is purely linear in the quantum fields. It is known that such
a kind of breaking does not get renormalized by the quantum corrections and
does not spoil the usefulness of the corresponding Ward identity \cite{book}%
. The latter turns out not only to be free from anomalies at the quantum
level, but it plays a crucial role in order to establish the ultraviolet
finiteness of the topological models \cite{dgs,gms,br}.

Let us also mention that the vector susy Ward operator can be introduced on
a more abstract geometrical way for a large class of gauge models \cite
{sslt,wos,womss,bss,cqs}, independently from the fact that it is or not
related to a (linearly broken) symmetry of the action. In this case the
vector Ward operator plays the role of an algebraic operator which, thanks
to the fact that it decomposes the space-time derivative as a BRST
anticommutator, turns out to be very useful in order to solve the descent
equations associated to the BRST cohomology classes for the anomalies and
the invariant counterterms. In addition, it allows to encode all the
relevant informations (BRST\ transformations of the fields, BRST\ cohomology
classes, solutions of the descent equations,...) into a unique equation
which takes the suggestive form of a generalized zero curvature condition 
\cite{zcurv,sdeseq}.

All these properties, if on the one hand make the vector susy quite
interesting, on the other hand motivate further investigations about its
origin. For instance, in the case of the Witten four dimensional topological
Yang-Mills theory one may wonder about the possibility of performing a twist 
\cite{tym} of the generators of the corresponding N=2 supersymmetric
Yang-Mills theory in order to obtain the vector susy. The situation is less
clear for other topological models, especially for those belonging to the
Schwartz class which are not manifestly related to an extended
supersymmetric algebra. For these models the vector susy has been introduced
essentially by hand \cite{brt,dgs,gms} and later on has been related to the
existence of a conserved current related to the fact that the
energy-momentum tensor of a topological theory can be expressed as a pure
BRST variation \cite{ms,book}.

Moreover, a general set up accounting for all the features displayed by the
vector susy has not yet been completely worked out. This is the aim of this
paper,\textit{\ i.e.} to provide a purely algebraic cohomological
characterization of the existence of the vector susy and of the related
linear classical breaking term. We shall also see that it is precisely the
requirement of linearity in the quantum fields of the breaking term which
selects a particular set of gauge parameters, clarifying thus the
relationship between the vector susy and the gauge fixing condition.

In particular, we shall be able to prove that the existence of the
topological vector susy turns out to be deeply related to a vector BRST
invariant antifield dependent cocycle with ghost number -1. The existence or
not of a vector supersymmetry depends purely on the fact that such an
antifield cocycle is cohomological trivial or not. When it is, the vector
susy Ward identity is present and turns out to be necessarily accompanied by
a breaking term linear in the quantum fields. On the other hand, when such
an antifield cocycle in not BRST\ exact, the vector susy cannot be
established and we are left with an example of a nontrivial antifield
dependent BRST cohomology class.

This algebraic framework is very related to the cohomological reformulation
of the Noether theorem given by M. Henneaux et al. \cite{mh}. We shall see
in fact that the aforementioned vector cocycle can be related to a set of
currents among which one can identify the BRST invariant energy-momentum
tensor, showing then that this antifield cocycle is related to the
Poincar\'{e} transformations, according to the analysis of Henneaux et al. 
\cite{mh}. Let us also mention that, recently, this vector antifield cocycle
has been considered in connection with the problem of including into a
unique extended Slavnov-Taylor identity additional global invariances of the
action \cite{glob}.

The paper is organized as follows. In the Sect.2 we introduce the general
algebraic set up, we present the relevant properties of the vector antifield
cocycle and we discuss its relation with the BRST invariant energy-momentum
tensor. In the Sect.3 the pure Yang-Mills model is considered. We shall
prove that in this case the vector cocycle is not trivial, identifying then
a BRST cohomology class. The Sect.4 is devoted to a detailed analysis of
several examples of topological models. It will be proven that in these
cases the BRST triviality of the vector cocycle is at the origin of the
existence of the linearly broken topological vector susy Ward identities.
Finally, in the Sect.5 we will present another example of a linearly broken
Ward identity whose contact terms are generated, in complete analogy with
the case of the topological models, by the BRST exactness of an antifield
dependent cocycle possessing a group index. Such a Ward identity, known as
the Landau ghost equation \cite{bps}, turns out to be related to the rigid
gauge invariance.

\section{General Notations and the Vector Cocycle}

In order to present the general algebraic set up let us begin by fixing the
notations. We shall work in a flat $D$-dimensional space-time equipped with
a set of fields generically denoted by $\left\{ \varphi ^i\right\} $, $i$
labelling the different kinds of fields needed in order to properly quantize
the model, \textit{i.e.} gauge fields, ghosts, ghosts for ghosts, etc...
Following the standard quantization procedure we introduce for each field $%
\varphi ^i$ of ghost number $\mathcal{N}_{\varphi ^i\text{ }}$ and dimension 
$d_{\varphi ^i}$, the corresponding antifield $\varphi ^{i*}$ with ghost
number $-\left( 1+\mathcal{N}_{\varphi ^i\text{ }}\right) $ and dimension $%
\left( D-d_{\varphi ^i}\right) $. We shall also assume that the set of
fields $\left\{ \varphi ^i\right\} $ do not explicitely contains the
antighosts and their corresponding Lagrange multipliers which, being grouped
in BRST doublets, do not contribute to the BRST cohomology \cite{ymcoh}.
Accordingly, some of the antifields $\varphi ^{i*}$ have to be understood as
shifted\textit{\ }antifields \cite{book} which already take into account the
antighosts.

In order to define the model it remains now to introduce the classical gauge
fixed \textit{reduced }\footnote{%
The name \textit{reduced }action is used to denote the part of the complete
gauge fixed action which does not depend from the Lagrangian multipliers and
which depends from the antighosts only through the shifted antifields. As it
will be discussed in the next Sections, the dependence of the complete
action from the Lagrangian multipliers as well as from the gauge parameters
present in the gauge fixing condition will be determined by requiring that
the breaking term associated to the topological vector susy Ward identity is
at most linear in the quantum fields.} action \cite{book} $\Sigma \left(
\varphi ,\varphi ^{*}\right) $. $\;$This is done by requiring that $\Sigma $
is$\;$power counting renormalizable and that it is the solution of the
classical homogeneous Slavnov-Taylor (or master equation) identity \cite{bv} 
\begin{equation}
\int d^Dx\frac{\delta \Sigma }{\delta \varphi ^i}\frac{\delta \Sigma }{%
\delta \varphi ^{i*}}\;=\;\frac 12\mathcal{B}_\Sigma \Sigma \;=\;0\;\;,
\label{slav-tayl}
\end{equation}
where $\mathcal{B}_\Sigma \;$denotes the nilpotent linearized Slavnov-Taylor
operator

\begin{equation}
\mathcal{B}_\Sigma \;=\int d^Dx\,\left( \frac{\delta \Sigma }{\delta \varphi
^i}\frac \delta {\delta \varphi ^{i*}}\;+\;\frac{\delta \Sigma }{\delta
\varphi ^{i*}}\frac \delta {\delta \varphi ^i}\right) \;\;,
\label{lin-slav-tayl}
\end{equation}

\begin{equation}
\mathcal{B}_\Sigma \;\mathcal{B}_\Sigma \;=\;0\;\;.  \label{nilp}
\end{equation}
As usual, the classical action $\Sigma \left( \varphi ,\varphi ^{*}\right) $
will be assumed to be invariant under the space-time translations, \textit{%
i.e.}

\begin{equation}
\mathcal{P}_\mu \Sigma \;=\;\int d^Dx\;\left( \partial _\mu \varphi ^i\frac{%
\delta \Sigma }{\delta \varphi ^i}\;+\;\partial _\mu \varphi ^{i*}\frac{%
\delta \Sigma }{\delta \varphi ^{i*}}\right) \;=\;0\;\;.
\label{translations}
\end{equation}
The classical Slavnov-Taylor identity $\left( \text{\ref{slav-tayl}}\right) $%
\ and the translation invariance $\left( \text{\ref{translations}}\right) $
will be taken therefore as the basic starting points for the
characterization of the classical action $\Sigma \left( \varphi ,\varphi
^{*}\right) $ and for the algebraic analysis which will be carried out in
the next sections. Let us also remark that the requirement of the
translation invariance, as expressed by the equation $\left( \text{\ref
{translations}}\right) $, does not imply any further restriction on $\Sigma $
than those that are tacetely assumed in any local field theory.

Let us introduce now the following integrated local polynomial, linear in
the antifields $\varphi ^{i*}$, of ghost number $-1$, dimension $\left(
D+1\right) $, and of the vector type

\begin{equation}
\Omega _\nu ^{-1}=\;\int d^Dx\;\left[ \omega _\nu ^{-1}\right]
_{D+1}\;\equiv \int d^Dx\;\left( -\right) ^{\left( 1+\mathcal{N}_{\varphi ^i%
\text{ }}\right) }\varphi ^i\;\partial _\nu \varphi ^{i*}\;\;.
\label{vec-cocycle}
\end{equation}
It is easily proven that the above expression is $\mathcal{B}_\Sigma -$%
invariant. In fact, one has

\begin{equation}
\mathcal{B}_\Sigma \;\Omega _\nu ^{-1}\;=\int d^Dx\;\left( \partial _\nu
\varphi ^i\frac{\delta \Sigma }{\delta \varphi ^i}\;+\;\partial _\nu \varphi
^{i*}\frac{\delta \Sigma }{\delta \varphi ^{i*}}\right) =\mathcal{P}_\nu
\Sigma \;=0\;\;,  \label{Om-inv}
\end{equation}
due to the translation invariance of the calssical action $\Sigma \left(
\varphi ,\varphi ^{*}\right) $. Having proven the invariance of $\Omega _\nu
^{-1}$ we have now to establish if it eventually identifies a cohomology
class of the operator $\mathcal{B}_\Sigma $. We have in fact the following
two possibilities, namely

\vspace{1cm}

$1)$ $\Omega _\nu ^{-1}$ is $\mathcal{B}_\Sigma -$exact, \textit{i.e.}

\begin{equation}
\Omega _\nu ^{-1}=\mathcal{B}_\Sigma \Xi _\nu ^{-2}\;\;,\;\;
\label{triviality}
\end{equation}
$\Xi _\nu ^{-2}\;$being an integrated local polynomial of ghost number $-2$
and dimension $\left( D+1\right) $.

\vspace{1cm}

$2)$ $\Omega _\nu ^{-1}$ belongs to the integrated cohomology of $\mathcal{B}%
_\Sigma $,

\begin{equation}
\Omega _\nu ^{-1}\neq \mathcal{B}_\Sigma \Xi _\nu ^{-2}\;.  \label{nontriv}
\end{equation}

The detailed analysis of these two possibilities will be the main subject of
the next Sections. Let us limit here to underline that while the first
possibility $\left( \ref{triviality}\right) $ turns out to be a feature of
the topological theories, the second one $\left( \ref{nontriv}\right) $ is
typical of a Yang-Mills theory and, more generally, of any model possessing
an invariant energy-momentum tensor $T_{\mu \nu }$ which cannot be written
as a pure $\mathcal{B}_\Sigma -$variation, \textit{i.e. }

\begin{equation}
\mathcal{B}_\Sigma T_{\mu \nu }=0\;\;,\;\;T_{\mu \nu }\neq \mathcal{B}%
_\Sigma \Lambda _{\mu \nu }\;\;.  \label{nontriv-en}
\end{equation}
In this case $\Omega _\nu ^{-1}$ provides an example of a nontrivial
antifield dependent cocycle of the operator $\mathcal{B}_\Sigma $, giving
thus an explicit realization of the general results of \cite{mh}.

Concerning now the first possibility$\;\left( \ref{triviality}\right) $, it
is worthwhile to recall that one of the peculiar property of the topological
models is precisely that of having a nonphysical energy momentum tensor \cite
{tym,bbrt}, \textit{i.e. }the energy momentum tensor of a topological field
theory $\;T_{\mu \nu }^{top}$ can be written as a pure $\mathcal{B}_\Sigma -$%
variation

\begin{equation}
T_{\mu \nu }^{top}=\mathcal{B}_\Sigma \Lambda _{\mu \nu }\;\;,
\label{triv-en}
\end{equation}
for some local $\Lambda _{\mu \nu }$, yielding thus a more direct indication
of the $\mathcal{B}_\Sigma -$exactness of the vector cocycle $\Omega _\nu
^{-1}$.

For a better understanding of the relation between $\Omega _\nu ^{-1}\;$and
the energy-momentum tensor, let us write down the descent equations \cite
{ymcoh,mh,book} corresponding to the integrated invariance condition $\left( 
\ref{Om-inv}\right) $, \textit{i.e.} 
\begin{eqnarray}
\mathcal{B}_{\Sigma \;}\left[ \omega _\nu ^{-1}\right] _{D+1}\;
&=&\;\partial ^{\mu _1}\left[ \omega _{\mu _1\nu }^0\right] _D\;\;,
\label{desc-equ} \\
\mathcal{B}_{\Sigma \;}\left[ \omega _{\mu _1\nu }^0\right] _D\;
&=&\;\partial ^{\mu _2}\left[ \omega _{\left[ \mu _1\mu _2\right] \nu
}^1\right] _{D-1}\;\;,  \nonumber \\
\mathcal{B}_{\Sigma \;}\left[ \omega _{\left[ \mu _1\mu _2\right] \nu
}^1\right] _{D-1}\; &=&\;\partial ^{\mu _3}\left[ \omega _{\left[ \mu _1\mu
_2\mu _3\right] \nu }^2\right] _{D-2}\;\;,  \nonumber \\
&&..........\;\;, \\
&&..........\;\;,  \nonumber \\
\mathcal{B}_{\Sigma \;}\left[ \omega _{\left[ \mu _1\mu _2...\mu
_{D-1}\right] \nu }^{D-2}\right] _2\; &=&\;\partial ^{\mu _D}\left[ \omega
_{\left[ \mu _1\mu _{2...}\mu _{D-1}\mu _D\right] \nu }^{D-1}\right] _1\;, 
\nonumber \\
\mathcal{B}_{\Sigma \;}\left[ \omega _{\left[ \mu _1\mu _{2...}\mu _{D-1}\mu
_D\right] \nu }^{D-1}\right] _1\; &=&0\;\;,  \nonumber
\end{eqnarray}
where the $\left[ \omega _{\left[ \mu _1\mu _{2...}\mu _j\right] \nu
}^{j-1}\right] _{D-j+1}$\ with $\left( j=0,...,D\right) $ are local currents
of ghost number $\left( j-1\right) $, antisymmetric in the lower indices$%
\;\left( \mu _{1,}\mu _{2,...,}\mu _j\right) $,$\;$and of dimension $\left(
D-j+1\right) $. The usefulness of working with the system $\left( \ref
{desc-equ}\right) \;$is due to the fact that these equations relate the
local cocycle $\left[ \omega _\nu ^{-1}\right] _{D+1}$ with currents of
lower dimension which can provide a more easy and transparent interpretation
of the physical meaning of $\Omega _\nu ^{-1}$. To this purpose it should be
remarked that the cocycle $\left[ \omega _{\mu _1\nu }^0\right] _D\;$%
entering the second equation of the system $\left( \ref{desc-equ}\right) \;$%
has the same quantum numbers of the energy momentum tensor, being of
dimension $D$, of ghost number $0$, and possessing two Lorentz indices. In
particular, from the general results on the BRST cohomology \cite{ymcoh} it
follows that, independently from the fact that the cohomology of the
operator $\mathcal{B}_{\Sigma \;}$is empty or not in the sectors with
dimension lower than $D$, the existence of a nontrivial local cohomology in
the sector of dimension $D$ and with two free Lorentz indices necessarily
implies the nontriviality of the upper level of dimension $D+1$. It becomes
apparent thus that the nontriviality of the BRST invariant energy-momentum
tensor is deeply related to the existence of a nontrivial integrated
cohomology in the sector of dimension $D+1\;$and negative ghost-number.

In fact, as we shall see explicitily in the case of the Yang-Mills theory,
the current$\;\left[ \omega _{\mu _1\nu }^0\right] _D$ turns out to be
precisely the improved BRST invariant energy-momentum tensor. On the other
hand, the use of the descent equations $\left( \ref{desc-equ}\right) $
provides a simple demonstration of the triviality of the vector cocycle $%
\Omega _\nu ^{-1}$ in the case of the topological theories. This is actually
due to the fact that the field content of the topological models gives rise
to BRST cohomology classes which are empty in the various sectors appearing
in the descent equations $\left( \ref{desc-equ}\right) $, implying in
particular that the corresponding energy momentum tensor is BRST trivial,
according to eq.$\left( \ref{triv-en}\right) $. Moreover, as we shall see in
detail later on, the right hand side of the equation $\left( \ref{triviality}%
\right) $ expressing the BRST triviality of $\Omega _\nu ^{-1}$ will provide
the contact terms of the vector susy Ward identity of the topological
theories. In this latter case we shall also check that the left hand side of
eq.$\left( \ref{triviality}\right) $ reduces to a classical breaking, 
\textit{i.e. }to a breaking purely linear in the quantum fields, showing
then that the topological vector susy Ward identity is always linearly
broken.

\section{Pure Yang-Mills Theory}

As a first application of the general algebraic set up discussed in the
previous Section, let us now prove that in the case of pure Yang-Mills
action the vector cocycle $\Omega _\nu ^{-1}$ cannot be written as an exact $%
\mathcal{B}_\Sigma -$term. Let us begin by considering the complete fully
quantized gauge fixed Yang-Mills action which, choosing a Feynman gauge,
reads

\begin{eqnarray}
\mathcal{S} &=&\int d^4x\;\left( -\frac 1{4g^2}F_{\mu \nu }^aF^{a\mu \nu }\;+%
\text{ }b^a\partial A^a+\frac \alpha 2b^ab^a\;+\;\partial ^\mu \overline{c}%
^a(D_\mu c)^a\right)  \label{y-m-action} \\
&&+\int d^4x\;\hat{A}_\mu ^{a*}(D^\mu c)^a-\frac 12\;f_{abc}C^{*a}c^bc^c\;, 
\nonumber
\end{eqnarray}
where $\left( c,\overline{c},b\right) \;$denote respectively the ghost, the
antighost and the Lagrange multiplier fields and

\begin{equation}
(D_\mu c)^a=\;\partial _\mu c^a\;+f_{\;bc}^a\;A_\mu ^bc^c\;\;,
\label{cov-der}
\end{equation}
is the covariant derivative with $f_{abc}$ the totally antisymmetric
structure constants of a compact semisimple Lie group $G$. The two
antifields $\left( \hat{A}^{*},C^{*}\right) \;$are introduced in order to
properly define the nonlinear BRST transformations of the gauge field $A$
and of the Faddeev-Popov ghost $c$. The quantum numbers, \textit{i.e.} the
dimensions and the ghost numbers, of all the fields and antifields are
assigned as in the following table 
\begin{table}[tbh]
\centering
\begin{tabular}{|c|c|c|c|c|c|c|}
\hline
& $A$ & ${\hat{A}}^{*}$ & $c$ & $C^{*}$ & $\bar{c}$ & $b$ \\ \hline
$dim$ & $1$ & $3$ & $0$ & $4$ & $2$ & $2$ \\ 
$gh-numb$ & $0$ & $-1$ & $1$ & $-2$ & $-1$ & $0$ \\ \hline
\end{tabular}
\caption[t1]{dimension and ghost number}
\label{ym-table}
\end{table}
\\As it is well known \cite{book}, the complete action $\mathcal{S}$ $\;$is
characterized by the classical Slavnov-Taylor identity, expressing the
invariance of $\left( \ref{y-m-action}\right) $ under the BRST
transformations, \textit{i.e.}

\begin{equation}
\int d^4x\left( \frac{\delta \mathcal{S}}{\delta A_\mu ^a}\frac{\delta 
\mathcal{S}}{\delta \hat{A}^{*a\mu }}+\frac{\delta \mathcal{S}}{\delta c^a}%
\frac{\delta \mathcal{S}}{\delta C^{*a}}+b^a\frac{\delta \mathcal{S}}{\delta 
\overline{c}^a}\right) \;=\;0\;\;,  \label{y-m-slav-tayl}
\end{equation}
and by the linear gauge fixing condition \cite{book}

\begin{equation}
\frac{\delta \mathcal{S}}{\delta b^a}\;=\;\partial A^a+\alpha b^a\;\;.
\label{feynm-gauge-fix}
\end{equation}
A further identity, the antighost equation \cite{book}

\begin{equation}
\frac{\delta \mathcal{S}}{\delta \overline{c}^a}\;+\;\partial ^\mu \frac{%
\delta \mathcal{S}}{\delta \hat{A}^{*a\mu }}\;=\;0\;\;  \label{ant-ghost-eq}
\end{equation}
follows from the gauge fixing condition $\left( \ref{feynm-gauge-fix}\right) 
$ and from the Slavnov-Taylor identity $\left( \ref{y-m-slav-tayl}\right) $,
implying that the antighost $\overline{c}$ and the antifield $\hat{A}^{*}$
can enter only through the shifted\textit{\ }antifield $A^{*a\mu }\;$of
ghost number -1 and dimension 3

\begin{equation}
A^{*a\mu }\;=\;\hat{A}^{*a\mu }\;+\;\partial ^\mu \overline{c}^a\;\;.
\label{shifted-antif}
\end{equation}
Introducing now the reduced Yang-Mills action $\Sigma \left(
A,A^{*},c,C^{*}\right) $ defined by the identities $\left( \ref
{feynm-gauge-fix}\right) $ and $\left( \ref{ant-ghost-eq}\right) $

\begin{equation}
\mathcal{S}\;=\;\Sigma \;+\int d^4x\;\left( \text{ }b^a\partial A^a+\frac
\alpha 2b^ab^a\right) \;\;\;,  \label{def-red-action}
\end{equation}
\textit{i.e.}

\begin{equation}
\Sigma \;=\int d^4x\;\left( -\frac 1{4g^2}F_{\mu \nu }^aF^{a\mu \nu
}\;+\;A^{*a\mu }(D_\mu c)^a-\frac 12\;f_{abc}C^{*a}c^bc^c\;\right) \;\;,
\label{red-ym-action}
\end{equation}
it is easily verified that $\Sigma $ obeys the homogeneous Slavnov-Taylor
identity 
\begin{equation}
\int d^4x\left( \frac{\delta \Sigma }{\delta A_\mu ^a}\frac{\delta \Sigma }{%
\delta A^{*a\mu }}\;+\frac{\delta \Sigma }{\delta c^a}\frac{\delta \Sigma }{%
\delta C^{*a}}\;\right) =\;\frac 12\mathcal{B}_\Sigma \Sigma \;=\;0\;\;,
\label{y-m-hom-st}
\end{equation}
with

\begin{equation}
\mathcal{B}_\Sigma \;=\;\int d^4x\left( \frac{\delta \Sigma }{\delta A_\mu ^a%
}\frac \delta {\delta A^{*a\mu }}+\frac{\delta \Sigma }{\delta A^{*a\mu }}%
\frac \delta {\delta A^{a\mu }}+\frac{\delta \Sigma }{\delta c^a}\frac
\delta {\delta C^{*a}}\;+\frac{\delta \Sigma }{\delta C^{*a}}\frac \delta
{\delta c^a}\right) \;\;,  \label{y-m-linear}
\end{equation}
and

\begin{equation}
\mathcal{B}_\Sigma \mathcal{B}_\Sigma \;=\;0\;\;.  \label{y-m-linear-nilp}
\end{equation}
For the vector cocycle $\Omega _\nu ^{-1}$ of eq. $\left( \ref{vec-cocycle}%
\right) $ we have now

\begin{equation}
\Omega _\nu ^{-1}=\;\int d^4x\;\;\left[ \omega _\nu ^{-1}\right] _5\;\equiv
\;\int d^4x\;\left( c^a\partial _\nu C^{*a}-A_\mu ^a\partial _\nu A^{*a\mu
}\;\right) \;,  \label{ym-vect-cocycle}
\end{equation}
which, according to the equation $\left( \ref{Om-inv}\right) $, turns out to
be $\mathcal{B}_\Sigma -$invariant 
\begin{equation}
\mathcal{B}_\Sigma \;\Omega _\nu ^{-1}=\;\mathcal{P}_\nu \Sigma \;=0\;\;.
\label{ym-Om-inv}
\end{equation}
We are now ready to prove that in the case of Yang-Mills theory the vector
cocycle $\Omega _\nu ^{-1}$ of eq.$\left( \ref{ym-vect-cocycle}\right) $ is
nontrivial. Let us proceed by assuming the converse, \textit{i.e.} let us
suppose that $\Omega _\nu ^{-1}$ can be written as an exact $\mathcal{B}%
_\Sigma -$term 
\begin{equation}
\Omega _\nu ^{-1}=\mathcal{B}_\Sigma \Xi _\nu ^{-2}\;\;,\;\;
\label{ym-triviality}
\end{equation}
for some integrated local polynomial $\Xi _\nu ^{-2}\;$of ghost number $-2$
and dimension $5$. From the Table 1 it follows that the most general form
for $\Xi _\nu ^{-2}$ is given by

\begin{equation}
\Xi _\nu ^{-2}\;=\beta \int d^4x\;C^{*a}A_\nu ^a\;\;,  \label{ym-exactness}
\end{equation}
where $\beta $ is an arbitrary free parameter which has to be fixed by the
exactness condition $\left( \ref{ym-triviality}\right) $. Thus, from eq.$%
\left( \ref{ym-triviality}\right) $, we should have the algebraic equality

\begin{equation}
\int d^4x\left( c^a\partial _\nu C^{*a}-A_\mu ^a\partial _\nu A^{*a\mu
}\right) =\beta \int d^4x\left( C^{*a}\partial _\nu c^a+A_\nu ^a\left( D_\mu
A^{*\mu }\right) ^a\;\right) \;\;.  \label{ym-eq}
\end{equation}
However it is almost immediate to check that the above equation has no
solution for $\beta $, showing then that $\Omega _\nu ^{-1}$cannot be
written as a pure $\mathcal{B}_\Sigma -$variation,

\begin{equation}
\Omega _\nu ^{-1}\neq \mathcal{B}_\Sigma \Xi _\nu ^{-2}\;\;.
\label{ym-nontriviality}
\end{equation}
Therefore, in the case of Yang-Mills theory the vector cocycle $\Omega _\nu
^{-1}$ identifies a cohomology class of the operator $\mathcal{B}_\Sigma \;$%
in the sector of the integrated local polynomials of ghost number -1 and
dimension 5, providing thus an explicit example of an antifield dependent
cohomology class of $\mathcal{B}_\Sigma $.

For a better understanding of the nontriviality of the vector cocycle $%
\Omega _\nu ^{-1}$ it remains now to discuss its relation with the
energy-momentum tensor of the model. To this purpose we analyse the descent
equations corresponding to the local polynomial $\left[ \omega _\nu
^{-1}\right] _5$, \textit{i.e. }

\begin{eqnarray}
\mathcal{B}_{\Sigma \;}\left[ \omega _\nu ^{-1}\right] _5\; &=&\;\partial
^{\mu _1}\left[ \omega _{\mu _1\nu }^0\right] _4\;\;,  \label{ym-desc-equ} \\
\mathcal{B}_{\Sigma \;}\left[ \omega _{\mu _1\nu }^0\right] _4\;
&=&\;\partial ^{\mu _2}\left[ \omega _{\left[ \mu _1\mu _2\right] \nu
}^1\right] _3\;\;,  \nonumber \\
\mathcal{B}_{\Sigma \;}\left[ \omega _{\left[ \mu _1\mu _2\right] \nu
}^1\right] _3\; &=&\;\partial ^{\mu _3}\left[ \omega _{\left[ \mu _1\mu
_2\mu _3\right] \nu }^2\right] _2\;\;,  \nonumber \\
\mathcal{B}_{\Sigma \;}\left[ \omega _{\left[ \mu _1\mu _2\mu _3\right] \nu
}^2\right] _2\; &=&\;\partial ^{\mu _4}\left[ \omega _{\left[ \mu _1\mu
_2\mu _3\mu _4\right] \nu }^3\right] _1\;,  \nonumber \\
\mathcal{B}_{\Sigma \;}\left[ \omega _{\left[ \mu _1\mu _2\mu _3\mu
_4\right] \nu }^3\right] _1\; &=&0\;\;.  \nonumber
\end{eqnarray}
\ After some straightforward algebraic manipulations, for the local currents 
$\left( \omega _{\left[ \mu _1..\mu _j\right] \nu }^{j-1},\;j=1,..4\right) $%
\ one finds

\begin{equation}
\left[ \omega _{\mu _1\nu }^0\right] _4\;=\frac 1{g^2}\;\left( F_{\mu
_1\sigma }^aF_\nu ^{a\;\sigma }\;-\;\frac 14g_{\mu _1\nu }F_{\rho \sigma
}^aF^{a\rho \sigma }\right) +\mathcal{B}_\Sigma \left( A_{\mu _1}^{*a}A_\nu
^a\right) \;\;,\;  \label{ym-en-mom}
\end{equation}

\begin{equation}
\left[ \omega _{\left[ \mu _1\mu _2\right] \nu }^1\right] _3\;=\varepsilon
_{\mu _1\mu _2}^{\;\;\;\;\;\;\;\mu _3\mu _4}\;\partial _{\mu _3}\left[ 
\tilde{\omega}_{\mu _4\nu }^1\right] _2\;\;,  \label{ot-1}
\end{equation}

\begin{equation}
\left[ \omega _{\left[ \mu _1\mu _2\mu _3\right] \nu }^2\right]
_2\;=\;\varepsilon _{\mu _1\mu _2\mu _3}^{\;\;\;\;\;\;\;\;\;\;\;\mu _4}%
\mathcal{B}_\Sigma \left[ \tilde{\omega}_{\mu _4\nu }^1\right]
_2\;\;+\varepsilon _{\mu _1\mu _2\mu _3}^{\;\;\;\;\;\;\;\;\;\;\;\mu
_4}\partial _{\mu _4}\left[ \tilde{\omega}_\nu ^2\right] _1\;\;,
\label{ot-2}
\end{equation}

\begin{equation}
\left[ \omega _{\left[ \mu _1\mu _2\mu _3\mu _4\right] \nu }^3\right]
_1\;=\varepsilon _{\mu _1\mu _2\mu _3\mu _4}^{\;\;\;\;\;\;\;}\mathcal{B}%
_\Sigma \left[ \tilde{\omega}_\nu ^2\right] _1\;\;,  \label{ot-3}
\end{equation}
with $\left[ \tilde{\omega}_{\mu _4\nu }^1\right] _2\;$and $\left[ \tilde{%
\omega}_\nu ^2\right] _1$ local arbitrary polynomials. From the above
expressions we observe that while the last three currents are trivial
solutions of the descent equations $\left( \ref{ym-desc-equ}\right) $, the
first one, \textit{i.e. }$\left[ \omega _{\mu _1\nu }^0\right] _4$,\textit{\ 
} yields the familiar expression of the BRST invariant improved
energy-momentum tensor which, as it is well known , belongs to the
cohomology of $\mathcal{B}_\Sigma $ \cite{ymcoh}. One sees thus that, as
already remarked in Sect.1, the existence of a nontrivial invariant
energy-momentum tensor is at the origin of the nontriviality of the vector
cocycle $\Omega _\nu ^{-1}$. Let us conclude this Section by remarking that
the\ exact term $\mathcal{B}_\Sigma \left( A_{\mu _1}^{*a}A_\nu ^a\right) $
which naturally appears in the right hand side of the equation $\left( \ref
{ym-en-mom}\right) $\ is needed in order to ensure the off-shell
conservation of the improved Yang-Mills energy-momentum tensor $T_{\mu \nu
}^{YM}=F_{\mu \sigma }^aF_\nu ^{a\;\sigma }-\frac 14g_{\mu \nu }F_{\rho
\sigma }^aF^{a\rho \sigma }$. Of course, the same conclusions hold if the
pure Yang-Mills action is supplemented with the introduction of matter
fields.

\section{The Case of the Topological Field Theories}

Having proven the nontriviality of $\Omega _\nu ^{-1}$ for Yang-Mills type
theories, let us now turn to analyse the vector cocycle $\left( \ref
{vec-cocycle}\right) $ in the context of the topological models. In this
case, as already mentioned in the Sect.2, it turns out that $\Omega _\nu
^{-1}$ is always BRST exact, \textit{i.e. }

\begin{equation}
\Omega _\nu ^{-1}=\mathcal{B}_\Sigma \Xi _\nu ^{-2}\;\;,\;\;
\label{top-triviality}
\end{equation}
for some integrated local field polynomial $\Xi _\nu ^{-2}$ with ghost
number $-2$. The reason of the exactness of $\Omega _\nu ^{-1}$ for the
topological theories relies on their field content which, as proven by
several authors \cite{osv,kk,bci,dmpw,gms,br,book}, does not allow for
nontrivial BRST cohomology classes with free Lorentz indices, as the ones
needed in order to have nontrivial solutions of the descent equations $%
\left( \ref{desc-equ}\right) $. As it is well known, the topological models
can be basically divided in two classes \cite{bbrt}, yielding respectively
the so called cohomological and Schwartz type theories, both having a BRST
trivial energy-momentum tensor. The models belonging to the first class are
identified by the fact that the gauge fixed classical action can be
expressed as a pure BRST\ variation\textit{\ }and that it is invariant under
the so called topological shift symmetry \cite{bbrt,osv,kk,bci,dmpw,bs,lp}.
In the second case the invariant action is not a pure BRST variation,
although it depends on the metric tensor only through the unphysical gauge
fixing term. Examples of theories belonging to the first class are given by
the Witten's topological Yang-Mills theory \cite{tym} and by the topological
sigma model \cite{tsm}. The three-dimensional Chern-Simons model \cite
{bbrt,brt,dgs} and the $BF$ systems \cite{bbrt,gms} provide examples of
Schwartz type topological theories.

Without entering into the details, let us limit here to mention that in the
case of the cohomological models the topological shift symmetry implies the
emptiness of the BRST\ cohomology\footnote{%
Let us remark here that, as discussed by R. Stora et al.\cite
{osv,kk,bci,dmpw}, the relevant cohomology for the topological theories of
the Witten's type is the so called equivariant cohomology. The latter is the
restriction of the BRST cohomology to gauge invariant local field
polynomials which do not depend from the Faddee-Popov ghost field $c$.
Contrary to the BRST cohomology, the equivariant cohomology is not empty and
turns out to provide a consistent definition of the Witten's observables.} 
\cite{bbrt,br,osv,kk,bci} and therefore the triviality of the vector cocycle 
$\Omega _\nu ^{-1}$. Concerning now the Schwartz type models, it can be
proven that the field content of these theories allows for nontrivial BRST\
cohomology classes \cite{dgs,gms,book}. Moreover, as observed in \cite{zcurv}%
, these models can be formulated in a pure geometrical way due to the fact
that all the fields (gauge fields, ghosts, ghosts for ghosts, etc...) can be
viewed as being the components of a generalized gauge connection obeying a
zero curvature condition. This structure, also called complete\textit{\ }%
ladder\textit{\ }structure, implies that all nontrivial BRST\ cohomology
classes can be identified with invariant polynomials built up with the
undifferentiated dimensionless scalar ghosts present in the model \cite
{dgs,gms,book}. In particular, from this result it follows that the BRST
cohomology classes entering the descent equations $\left( \ref{desc-equ}%
\right) $ are empty, meaning thus that also in the case of the Schwartz type
topological theories the vector cocycle $\Omega _\nu ^{-1}$ is BRST\ exact .

Although the equation $\left( \ref{top-triviality}\right) $\ could seem
empty of any interesting information because of its BRST\ exactness, it
turns out however that its content is far from being irrelevant. In fact, as
we shall see explicitely in the models considered below, the right hand side
of the eq.$\left( \ref{top-triviality}\right) $, due to the form $\left( \ref
{lin-slav-tayl}\right) $ of the operator $\mathcal{B}_\Sigma $, is easily
seen to provide the contact terms of a Ward identity. This identity, due to
the presence of the vector cocycle $\Omega _\nu ^{-1}$ in the left hand side
of $\left( \ref{top-triviality}\right) $, cannot express an exact invariance
of the theory. Instead, eq.$\left( \ref{top-triviality}\right) \;$will have
the meaning of a broken Ward identity. Moreover, it is not difficult to
convince oneself that such a breaking term is in fact a classical breaking, 
\textit{i.e. }a breaking which is linear in the quantum fields. Therefore it
does not get renormalized by the radiative corrections and does not spoil
the usefulness of the corresponding broken Ward identity. As one can easily
understand, the unavoidable existence of this classical breaking term is a
consequence of the fact that the general expression of the vector cocycle $%
\Omega _\nu ^{-1}$ given in eq.$\left( \ref{vec-cocycle}\right) $ is linear
in the quantum fields, apart possible quadratic terms coming from the fact
that some of the antifields $\varphi ^{i*}$ have been shifted in order to
take into account the antighosts. However we shall check that these
quadratic terms turn out to be handled in a very simple way, being
reinterpreted as pure contact terms by means of an appropriate choice of the
nonphysical gauge parameters present in the gauge condition. This means that
the requirement of having breaking terms at most linear in the quantum
fields will fix the gauge parameters, implying therefore that the
corresponding Ward identity is consistent with the Quantum Action Principles 
\cite{qap} only for certain classes of gauge fixings. We shall give explicit
examples of this feature later on in the cases of the three dimensional
Chern-Simons model and of the Landau ghost equation derived in the Sect.5.

The equation $\left( \ref{top-triviality}\right) $ has thus the meaning of a
purely algebraic cohomological characterization of a Ward identity. In fact,
once one is able to prove that the vector cocycle $\Omega _\nu ^{-1}$ is
trivial, one knows authomatically that the equation $\left( \ref
{top-triviality}\right) $ has to be necessarily satisfied for some local
polynomial $\Xi _\nu ^{-2}$. The strategy to be followed now is almost
trivial. One writes down the most general expression for $\Xi _\nu ^{-2}$
compatible with the dimension of the space-time and with the field content
of the model under consideration. Such an expression will, in general,
depend on a set of free global coefficients. These coefficients are fixed by
demanding that eq.$\left( \ref{top-triviality}\right) $ is valid. Moreover,
recalling that the functional form of the operator $\mathcal{B}_\Sigma $
(see eq. $\left( \ref{lin-slav-tayl}\right) $) depends on the reduced action 
$\Sigma $, it is apparent to see that the coefficients of $\Xi _\nu ^{-2}$
as well as the contact terms of the corresponding Ward identity, are
uniquely determined by the form of the classical action $\Sigma $. This
means that, whenever the cocycle $\Omega _\nu ^{-1}$ is trivial, the
equation $\left( \ref{top-triviality}\right) $ tells us that the classical
action $\Sigma $, in addition to the Slavnov-Taylor identity, obeys a
further Ward identity whose contact terms are precisely given by the
equation $\left( \ref{top-triviality}\right) $, providing thus a simple
mechanism for an algebraic characterization of a new linearly broken
symmetry of the action. Of course such an additional Ward identity will have
the same quantum numbers of the vector cocycle $\Omega _\nu ^{-1}$, \textit{%
i.e. }it will carry a Lorentz index and will have a negative ghost number.
This vector identity, always present in the topological field theories, is
known in the litterature as the topological vector supersymmetry \cite
{brt,dgs,gms,br}. \textit{\ }

It is important at this point to spend some words on the possibility of
generalizing this purely algebraic mechanism in order to find other kinds of
unknown symmetries eventually present in the model. What seems to emerge
from the previous analysis is that the antifield dependent cocycles which
are relevant for the existence of a \textit{class of linearly broken} new
Ward identities are those which are linear in the fields $\left\{ \varphi
^i\right\} $ and in the antifields $\left\{ \varphi ^{i*}\right\} $ and that
are BRST\ exact. It is in fact this last property which when cast in the
form of the equation $\left( \ref{top-triviality}\right) $ allows us to
identify the contact terms of the corresponding Ward identity thanks to the
form of the linearized Slavnov-Taylor operator. It is worthwhile to remark
that an equation of the kind of $\left( \ref{top-triviality}\right) $
implies that the associate Ward identity is always broken, due to the
existence of a nonvanishing left hand side term. Moreover, being this term
linear in the quantum fields, the resulting breaking term is classical. In
the next chapter we will discuss another interesting example of a
classically linearly broken Ward identity, associated to the rigid gauge
invariance, whose contact terms can be characterized in a purely algebraic
way by means of an equation of the type of $\left( \ref{top-triviality}%
\right) $.

Let us focus, for the time being, on the discussion of the vector susy Ward
identity in the case of the topological field theories. In order to present
in an explicit way the previous algebraic mechanism, we shall analyse in
detail the examples of the three dimensional Chern-Simons gauge model, of
the four dimensional $BF$ model and of the so called two dimensional $b$-$c$
ghost system. The argument can be easily repeated and adapetd to other kinds
of topological models, such as the Witten's cohomological theories and the
higher dimensional $BF$ models.

\subsection{The three dimensional Chern-Simons model}

Using the same notations of the Sect.3, for the complete three dimensional
Chern-Simons action \cite{bbrt} quantized in the Landau gauge we have 
\begin{equation}
\mathcal{S}=\mathcal{S}_{cs}+\int d^3x\;\left( b^a\partial A^a+\partial ^\mu 
\overline{c}^a(D_\mu c)^a+\;\hat{A}_\mu ^{*}(D^\mu c)^a-\frac
12\;f_{abc}C^{*a}c^bc^c\right) \;,  \label{c-s-quant}
\end{equation}
where $\mathcal{S}_{cs}\;$is given by

\begin{equation}
\mathcal{S}_{cs}=-\frac k{4\pi }\int d^3x\;\varepsilon ^{\mu \nu \rho
}\left( A_\mu ^a\partial _\nu A_\rho ^a+\frac 13\;f_{abc}\text{ }A_\mu
^aA_\nu ^bA_\rho ^c\right) \;,  \label{c-s-action}
\end{equation}
$k$ identifying the inverse of the coupling constant. The fields and the
antifields have now the following quantum numbers 
\begin{table}[tbh]
\centering
\begin{tabular}{|c|c|c|c|c|c|c|}
\hline
& $A$ & ${\hat{A}}^{*}$ & $c$ & $C^{*}$ & $\bar{c}$ & $b$ \\ \hline
$dim$ & $1$ & $2$ & $0$ & $3$ & $1$ & $1$ \\ 
$gh-numb$ & $0$ & $-1$ & $1$ & $-2$ & $-1$ & $0$ \\ \hline
\end{tabular}
\caption[t2]{dimension and ghost number}
\label{cstable}
\end{table}
\\For the classical Slavnov-Taylor identity one has

\begin{equation}
\int d^3x\left( \frac{\delta \mathcal{S}}{\delta A_\mu ^a}\frac{\delta 
\mathcal{S}}{\delta \hat{A}^{*a\mu }}+\frac{\delta \mathcal{S}}{\delta c^a}%
\frac{\delta \mathcal{S}}{\delta C^{*a}}+b^a\frac{\delta \mathcal{S}}{\delta 
\overline{c}^a}\right) \;=\;0\;\;.  \label{c-s-slav-tayl}
\end{equation}
As before, making use of the Landau gauge condition

\begin{equation}
\frac{\delta \mathcal{S}}{\delta b^a}\;=\;\partial A^a\;\;,
\label{landau-gauge-fix}
\end{equation}
and of the antighost equation

\begin{equation}
\frac{\delta \mathcal{S}}{\delta \overline{c}^a}\;+\;\partial ^\mu \frac{%
\delta \mathcal{S}}{\delta \hat{A}^{*a\mu }}\;=\;0\;\;,
\label{c-s-ant-ghost-eq}
\end{equation}
for the reduced Chern-Simons action $\Sigma \left( A,A^{*},c,C^{*}\right) $

\begin{equation}
\mathcal{S}\;=\;\Sigma \;+\int d^3x\;\text{ }b^a\partial A^a\;\;,
\label{c-s-def-red-action}
\end{equation}
we obtain

\begin{equation}
\Sigma \;=\mathcal{S}_{cs}+\int d^3x\;\left( A_\mu ^{a*}(D^\mu c)^a-\frac
12\;f_{abc}C^{*a}c^bc^c\;\right) \;,  \label{red-cs-action}
\end{equation}
where, as usual, $A_\mu ^{a*}\;$is the shifted antifield

\begin{equation}
A^{*a\mu }\;=\;\hat{A}^{*a\mu }\;+\;\partial ^\mu \overline{c}^a\;\;.
\label{shifted-cs-antif}
\end{equation}
Finally, taking into account the eqs. $\left( \ref{landau-gauge-fix}\right) $%
, $\left( \ref{c-s-ant-ghost-eq}\right) $ and the expression $\left( \ref
{red-cs-action}\right) $, the Slavnov-Taylor identity $\left( \ref
{c-s-slav-tayl}\right) $ becomes

\begin{equation}
\int d^3x\left( \frac{\delta \Sigma }{\delta A_\mu ^a}\frac{\delta \Sigma }{%
\delta A^{*a\mu }}\;+\frac{\delta \Sigma }{\delta c^a}\frac{\delta \Sigma }{%
\delta C^{*a}}\right) =\;\frac 12\mathcal{B}_\Sigma \Sigma \;=\;0\;\;,
\label{c-s-hom-st}
\end{equation}
with

\begin{equation}
\mathcal{B}_\Sigma \;=\;\int d^3x\left( \frac{\delta \Sigma }{\delta A_\mu ^a%
}\frac \delta {\delta A^{*a\mu }}+\frac{\delta \Sigma }{\delta A^{*a\mu }}%
\frac \delta {\delta A^{a\mu }}+\frac{\delta \Sigma }{\delta c^a}\frac
\delta {\delta C^{*a}}\;+\frac{\delta \Sigma }{\delta C^{*a}}\frac \delta
{\delta c^a}\right) \;\;,  \label{c-s-linear}
\end{equation}
and

\begin{equation}
\mathcal{B}_\Sigma \mathcal{B}_\Sigma \;=\;0\;\;.  \label{c-s-linear-nilp}
\end{equation}
Repeating the same procedure done in the case of Yang-Mills theory, for the
vector cocycle $\Omega _\nu ^{-1}$ we can write

\begin{equation}
\Omega _\nu ^{-1}=\;\int d^3x\;\;\left[ \omega _\nu ^{-1}\right] _4\;\equiv
\;\int d^3x\;\left( c^a\partial _\nu C^{*a}-A_\mu ^a\partial _\nu A^{*a\mu
}\;\right) \;.  \label{cs-vect-cocycle}
\end{equation}
and 
\begin{equation}
\mathcal{B}_\Sigma \;\Omega _\nu ^{-1}=\mathcal{P}_\nu \Sigma \;=0\;\;.
\label{Om-inv1}
\end{equation}
In this case, as already mentioned, $\Omega _\nu ^{-1}$ turns out to be exact

\begin{equation}
\Omega _\nu ^{-1}=\mathcal{B}_\Sigma \Xi _\nu ^{-2}\;,  \label{cs-exact}
\end{equation}
where $\Xi _\nu ^{-2}$ is an integrated local polynomial of dimension 4 and
ghost number -2. From the Table $\ref{cstable}$ it follows that the most
general expression for $\Xi _\nu ^{-2}$ is given by

\begin{equation}
\Xi _\nu ^{-2}=\int d^3x\;\left( \gamma C^{*a}A_\nu ^a+\;\frac \beta
2\varepsilon _{\nu \sigma \tau }A^{*a\sigma }A^{*a\tau }\right) \;,
\label{cs-xi-cocycle}
\end{equation}
where $\gamma $ and $\beta $ are free arbitrary parameters. Using now the
expressions of the linearized Slavnov-Taylor operator $\left( \ref
{c-s-linear}\right) $ and of the reduced Chern-Simons action $\left( \ref
{red-cs-action}\right) $, for the right hand side of eq. $\left( \ref
{cs-exact}\right) $ we have

\begin{equation}
\mathcal{B}_\Sigma \Xi _\nu ^{-2}\;=\int d^3x\;\left( \gamma A_\nu ^a\frac{%
\delta \Sigma }{\delta c^a}+\gamma C^{*a}\frac{\delta \Sigma }{\delta
A^{*a\nu }}-\beta \varepsilon _{\nu \sigma \tau }A^{*a\sigma }\frac{\delta
\Sigma }{\delta A_\tau ^a}\right) \;.  \label{cs-xi-express}
\end{equation}
Comparing then both sides of eq.$\left( \ref{cs-exact}\right) $, for the
coeffiecients $\gamma $ and $\beta $ we get

\begin{equation}
\gamma =-1\;,\;\;\;\;\;\beta =\frac{2\pi }k\;,  \label{cs-coeff}
\end{equation}
so that eq.$\left( \ref{cs-exact}\right) $\ can be rewritten as

\begin{equation}
\int d^3x\;\left( A_\nu ^a\frac{\delta \Sigma }{\delta c^a}+C^{*a}\frac{%
\delta \Sigma }{\delta A^{*a\nu }}+\frac{2\pi }k\varepsilon _{\nu \sigma
\tau }A^{*a\sigma }\frac{\delta \Sigma }{\delta A_\tau ^a}+c^a\partial _\nu
C^{*a}-A_\mu ^a\partial _\nu A^{*a\mu }\right) =0\;.  \label{cs-red-ward-id}
\end{equation}
Finally, moving from the reduced action $\Sigma \;$to the complete action $%
\mathcal{S}$\ of eq. $\left( \ref{c-s-quant}\right) $ and making use of the
gauge condition $\left( \ref{landau-gauge-fix}\right) $, the identity $%
\left( \ref{cs-red-ward-id}\right) $ can be cast in the form of a linearly
broken Ward identity, namely

\begin{equation}
\mathcal{W}_\nu \mathcal{S=}\Delta _\nu ^{cl}\;,  \label{cs-susy}
\end{equation}
with

\begin{equation}
\mathcal{W}_\nu =\int d^3x\left( A_\nu ^a\frac \delta {\delta
c^a}+C^{*a}\frac \delta {\delta \hat{A}^{*a\nu }}+\frac{2\pi }k\varepsilon
_{\nu \sigma \tau }\left( \hat{A}^{*a\sigma }+\partial ^\sigma \overline{c}%
^a\right) \frac \delta {\delta A_\tau ^a}+\partial _\nu \overline{c}^a\frac
\delta {\delta b^a}\right) \;,  \label{cs-W-op}
\end{equation}
and

\begin{equation}
\Delta _\nu ^{cl}=\int d^3x\;\left( C^{*a}\partial _\nu c^a-\hat{A}^{*a\mu
}\partial _\nu A_\mu ^a\;-\frac{2\pi }k\varepsilon _{\nu \sigma \tau }\hat{A}%
^{*a\sigma }\partial ^\tau b^a\right) \;.  \label{cs-breaking}
\end{equation}
The equation $\left( \ref{cs-susy}\right) $ is easily recognized to be the
well known topological vector susy Ward identity of the three dimensional
Chern-Simons theory \cite{brt,dgs}. Let us conclude this section by
observing that in the case in which instead of a Landau gauge fixing
condition we had adopted a Feynman gauge (see eq.$\left( \ref
{feynm-gauge-fix}\right) $), the right hand side of the eq. $\left( \ref
{cs-susy}\right) $ would have been modified by the additional term

\begin{equation}
\alpha \int d^3x\;b^a\partial _\nu \overline{c}^a\;,  \label{quadr-break}
\end{equation}
which, being quadratic in the quantum fields, would have spoiled the
usefulness of the identity $\left( \ref{cs-susy}\right) $. We see therefore
that, as already remarked, the requirement that the breaking term $\Delta
_\nu ^{cl}$ is a classical breaking, \textit{i.e. }at most linear in the
quantum fields, forces the gauge parameter $\alpha $ to vanish, \textit{i.e. 
}$\alpha =0$, picking up thus the Landau gauge.

\subsection{The four dimensional BF model}

As the second example, we shall present the case of the four dimensional $BF$
model \cite{bbrt} whose classical invariant action is given by 
\begin{equation}
\mathcal{S}_{BF}=-\frac 14\int d^4x\;\varepsilon ^{\mu \nu \rho \sigma
}F_{\mu \nu }^aB_{\rho \sigma }^a\;.  \label{bf-action}
\end{equation}
The quantization of this model requires the Batalin-Vilkoviski procedure 
\cite{bv} due to the presence of ghosts for ghosts. We shall limit here only
to report the final result, reminding the reader to the numerous references 
\cite{bbrt} for the technical details. In particular, using the same
notations of ref. \cite{gms}, in order to gauge fix the invariant action $%
\left( \ref{bf-action}\right) $ we introduce a set of lagrangian multipliers 
$\left( b^a,h_\mu ^a,\omega ^a,\lambda ^a\right) $, a set of antighosts $%
\left( \overline{c}^a,\overline{\xi }_\mu ^a,\overline{\phi }^a,e^a\right) $%
, and a triple of ghosts $\left( c^a,\xi _\mu ^a,\phi ^a\right) $. For the
gauge fixing action $\mathcal{S}_{gf}$ we have 
\begin{equation}
\begin{array}{ll}
\mathcal{S}_{gf}\;=\displaystyle\int d^4x & \left( b^a\partial A^a-\partial
^\mu \overline{c}^a(D_\mu c)^a+h_\nu ^a\partial _\mu B^{a\mu \nu }+\omega
^a\partial \xi ^a+h^{a\mu }\partial _\mu e^a+\omega ^a\lambda ^a\right.  \\ 
& -\partial ^\mu \overline{\phi }^a\left( (D_\mu \phi )^a+f_{abc}c^b\xi _\mu
^c\right) +\displaystyle\frac 12f_{abc}\varepsilon ^{\mu \nu \rho \sigma
}(\partial _\mu \overline{\xi }_\nu ^a)\ (\partial _\rho \overline{\xi }%
_\sigma ^b)\phi ^c \\ 
& \left. -\partial ^\mu \overline{\xi }^{a\nu }\left( (D_\mu \xi _\nu
)^a-(D_\nu \xi _\mu )^a+f_{abc}B_{\mu \nu }^bc^c\right) -\lambda ^a\partial 
\overline{\xi }^a\;\right) \;\;.
\end{array}
\label{bf-gauge-fix}
\end{equation}
The ghost numbers and the dimensions of all the fields and ghosts are
assigned as follows

\begin{table}[tbh]
\centering
\begin{tabular}{|c|c|c|c|c|c|c|c|c|c|c|c|c|c|}
\hline
& $A$ & ${B}$ & $c$ & $\xi $ & $\phi $ & $\bar{c}$ & $\overline{\xi }$ & $%
\overline{\phi }$ & $e$ & $b$ & $h$ & $\omega $ & $\lambda $ \\ \hline
$dim$ & $1$ & $2$ & $0$ & $1$ & $0$ & $2$ & $1$ & $2$ & $2$ & $2$ & $1$ & $2$
& $2$ \\ 
$gh-numb$ & $0$ & $0$ & $1$ & $1$ & $2$ & $-1$ & $-1$ & $-2$ & $0$ & $0$ & $%
0 $ & $-1$ & $1$ \\ \hline
\end{tabular}
\caption[t2]{dimension and ghost number}
\label{bftable}
\end{table}
\noindent Introducing now a set of antifields $\left( \;\hat{A}^{*a\mu },%
\hat{B}^{*a\mu \nu },C^{*a},\phi ^{*a},\hat{\xi}^{*a\mu }\right) $
associated respectively to the fields $\left( A^{a\mu },B^{a\mu \nu
},c^a,\phi ^a,\xi ^{a\mu }\right) $, \textit{i.e.} 
\begin{equation}
\begin{array}{ll}
\mathcal{S}_{ext}=\displaystyle\int d^4x & \left( \displaystyle\frac 12\hat{B%
}^{*a\mu \nu }\left( (D_\nu \xi _\mu )^a-(D_\mu \xi _\nu )^a-f_{abc}B_{\mu
\nu }^bc^c+f_{abc}\varepsilon _{\mu \nu \rho \sigma }\partial ^\rho 
\overline{\xi }^{b\sigma }\phi ^c\right) \right. \\ 
& +\hat{\xi}^{*a\mu }\left( (D_\mu \phi )^a+f_{abc}c^b\xi _\mu ^c\right) +%
\displaystyle\frac 18f_{abc}\varepsilon _{\mu \nu \rho \sigma }\hat{B}%
^{*a\mu \nu }\hat{B}^{*b\rho \sigma }\phi ^c \\ 
& \left. -\hat{A}^{*a\mu }(D_\mu c)^a+\displaystyle\frac
12f_{abc}C^{*a}c^bc^c+f_{abc}\phi ^{*a}c^b\phi ^c\right) \;\;\;,
\end{array}
\label{bf-ext-fields}
\end{equation}

\begin{table}[tbh]
\centering
\begin{tabular}{|c|c|c|c|c|c|}
\hline
& $\hat{A}^{*}$ & $\hat{B}^{*}$ & $C^{*}$ & $\phi ^{*}$ & $\hat{\xi}^{*}$ \\ 
\hline
$dim$ & $3$ & $2$ & $4$ & $4$ & $3$ \\ 
$gh-numb$ & $-1$ & $-1$ & $-2$ & $-3$ & $-2$ \\ \hline
\end{tabular}
\caption[t2]{dimension and ghost number}
\label{bf-ext-table}
\end{table}
\noindent we have that the complete action $\mathcal{S}$

\begin{equation}
\mathcal{S=S}_{BF}+\mathcal{S}_{gf}+\mathcal{S}_{ext}\;\;,
\label{bf-comp-act}
\end{equation}
obeys the following Slavnov-Taylor identity

\begin{equation}
\begin{array}{ll}
\displaystyle\int d^4x & \left( \displaystyle\frac{\delta \mathcal{S}}{%
\delta A_\mu ^a}\displaystyle\frac{\delta \mathcal{S}}{\delta \hat{A}^{*a\mu
}}+\displaystyle\frac{\delta \mathcal{S}}{\delta c^a}\displaystyle\frac{%
\delta \mathcal{S}}{\delta C^{*a}}+\displaystyle\frac 12\displaystyle\frac{%
\delta \mathcal{S}}{\delta B_{\mu \nu }^a}\displaystyle\frac{\delta \mathcal{%
S}}{\delta \hat{B}^{*a\mu \nu }}+\displaystyle\frac{\delta \mathcal{S}}{%
\delta \phi ^a}\displaystyle\frac{\delta \mathcal{S}}{\delta \phi ^{*a}}%
\right. \\ 
& \left. +\displaystyle\frac{\delta \mathcal{S}}{\delta \xi _\mu ^a}%
\displaystyle\frac{\delta \mathcal{S}}{\delta \hat{\xi}^{*a\mu }}+h_\mu ^a%
\displaystyle\frac{\delta \mathcal{S}}{\delta \overline{\xi }_\mu ^a}+b^a%
\displaystyle\frac{\delta \mathcal{S}}{\delta \overline{c}^a}+\omega ^a%
\displaystyle\frac{\delta \mathcal{S}}{\delta \overline{\phi }^a}+\lambda ^a%
\displaystyle\frac{\delta \mathcal{S}}{\delta e^a}\right) \;\;=\;0\;.
\end{array}
\label{bf-slav-tayl}
\end{equation}
In order to define the reduced action for the $BF$ model let us write down
the gauge-fixing conditions 
\begin{equation}
\begin{array}{ll}
\displaystyle\frac{\delta \mathcal{S}}{\delta b^a}\;=\;\partial A^a\;\;, & %
\displaystyle\frac{\delta \mathcal{S}}{\delta h^{a\mu }}=\partial _\mu
e^a\;+\;\partial ^\nu B_{\nu \mu \;}^a,\;\;\; \\ 
&  \\ 
\displaystyle\frac{\delta \mathcal{S}}{\delta \omega ^a}=\;\lambda
^a\;+\;\partial \xi ^a\;, & \displaystyle\frac{\delta \mathcal{S}}{\delta
\lambda ^a}\;=\;-\partial \overline{\xi }^a\;-\;\omega ^a\;.
\end{array}
\label{bf-gauge-fix}
\end{equation}
Commuting now the above conditions with the Slavnov-Taylor identity $\left( 
\ref{bf-slav-tayl}\right) $ we get the antighost equations

\begin{equation}
\begin{array}{ll}
\displaystyle\frac{\delta \mathcal{S}}{\delta \overline{c}^a}\;+\;\partial
^\mu \displaystyle\frac{\delta \mathcal{S}}{\delta \hat{A}^{*a\mu }}\;=\;0\;,
& \displaystyle\frac{\delta \mathcal{S}}{\delta \overline{\phi }^a}%
\;-\;\partial ^\mu \displaystyle\frac{\delta \mathcal{S}}{\delta \hat{\xi}%
^{*a\mu }}\;=\;0\;\;, \\ 
&  \\ 
\displaystyle\frac{\delta \mathcal{S}}{\delta e^a}\;=\;-\partial h^a\;\;, & %
\displaystyle\frac{\delta \mathcal{S}}{\delta \overline{\xi }^{a\nu }}%
\;+\;\partial ^\mu \displaystyle\frac{\delta \mathcal{S}}{\delta \hat{B}%
^{*a\mu \nu }}\;=\;-\partial _\nu \lambda ^a\;\;,
\end{array}
\label{bf-antgh-eq}
\end{equation}
so that, introducing the following shifted antifields 
\begin{eqnarray}
A^{*a\mu } &=&\;\;\hat{A}^{*a\mu }+\partial ^\mu \overline{c}^a\;\;,\;\;\xi
^{*a\mu }=\hat{\xi}^{*a\mu }-\partial ^\mu \overline{\phi }^a\;,
\label{bf-shif-antf} \\
B^{*a\mu \nu } &=&\hat{B}^{*a\mu \nu }+\left( \partial ^\mu \overline{\xi }%
^{a\nu }-\partial ^\nu \overline{\xi }^{a\mu }\right) \;,  \nonumber
\end{eqnarray}
for the reduced $BF\;$action $\Sigma $ we get

\begin{equation}
\mathcal{S}=\Sigma +\displaystyle\int d^4x\left( b^a\partial A^a+h_\nu
^a\partial _\mu B^{a\mu \nu }+\omega ^a\partial \xi ^a+h^{a\mu }\partial
_\mu e^a+\omega ^a\lambda ^a+-\lambda ^a\partial \overline{\xi }^a\right) \;,
\label{bf-act-red}
\end{equation}
and 
\begin{equation}
\begin{array}{ll}
\Sigma =\displaystyle\int d^4x & \left( -\displaystyle\frac 14\;\varepsilon
^{\mu \nu \rho \sigma }F_{\mu \nu }^aB_{\rho \sigma }^a\right. \; \\ 
& +\displaystyle\frac 12B^{*a\mu \nu }\left( (D_\nu \xi _\mu )^a-(D_\mu \xi
_\nu )^a-f_{abc}B_{\mu \nu }^bc^c+f_{abc}\varepsilon _{\mu \nu \rho \sigma
}\partial ^\rho \overline{\xi }^{b\sigma }\phi ^c\right) \\ 
& +\xi ^{*a\mu }\left( (D_\mu \phi )^a+f_{abc}c^b\xi _\mu ^c\right) +%
\displaystyle\frac 18f_{abc}\varepsilon _{\mu \nu \rho \sigma }B^{*a\mu \nu
}B^{*b\rho \sigma }\phi ^c \\ 
& \left. -A^{*a\mu }(D_\mu c)^a+\displaystyle\frac
12f_{abc}C^{*a}c^bc^c+f_{abc}\phi ^{*a}c^b\phi ^c\right) \;.
\end{array}
\label{bf-red-action}
\end{equation}
As usual, the reduced action $\Sigma $ obeys the homogeneous Slavnov-Taylor

\begin{equation}
\begin{array}{ll}
\frac 12\mathcal{B}_\Sigma \Sigma =0=\displaystyle\int d^4x & \left( %
\displaystyle\frac{\delta \Sigma }{\delta A_\mu ^a}\displaystyle\frac{\delta
\Sigma }{\delta A^{*a\mu }}+\displaystyle\frac{\delta \Sigma }{\delta c^a}%
\displaystyle\frac{\delta \Sigma }{\delta C^{*a}}+\displaystyle\frac 12%
\displaystyle\frac{\delta \Sigma }{\delta B_{\mu \nu }^a}\displaystyle\frac{%
\delta \Sigma }{\delta B^{*a\mu \nu }}\right. \\ 
&  \\ 
& \left. +\displaystyle\frac{\delta \Sigma }{\delta \phi ^a}\displaystyle%
\frac{\delta \Sigma }{\delta \phi ^{*a}}+\displaystyle\frac{\delta \Sigma }{%
\delta \xi _\mu ^a}\displaystyle\frac{\delta \Sigma }{\delta \xi ^{*a\mu }}%
\right) \;,
\end{array}
\label{bf-hom-slv-tayl}
\end{equation}
with

\begin{equation}
\begin{array}{ll}
\mathcal{B}_\Sigma =\displaystyle\int d^4x & \left( \displaystyle\frac{%
\delta \Sigma }{\delta A_\mu ^a}\displaystyle\frac \delta {\delta A^{*a\mu
}}+\displaystyle\frac{\delta \Sigma }{\delta A^{*a\mu }}\displaystyle\frac
\delta {\delta A_\mu ^a}+\displaystyle\frac{\delta \Sigma }{\delta c^a}%
\displaystyle\frac \delta {\delta C^{*a}}+\displaystyle\frac{\delta \Sigma }{%
\delta C^{*a}}\displaystyle\frac \delta {\delta c^a}\right. \\ 
&  \\ 
& +\displaystyle\frac 12\displaystyle\frac{\delta \Sigma }{\delta B_{\mu \nu
}^a}\displaystyle\frac \delta {\delta B^{*a\mu \nu }}+\displaystyle\frac 12%
\displaystyle\frac{\delta \Sigma }{\delta B^{*a\mu \nu }}\displaystyle\frac
\delta {\delta B_{\mu \nu }^a}+\displaystyle\frac{\delta \Sigma }{\delta
\phi ^a}\displaystyle\frac \delta {\delta \phi ^{*a}} \\ 
&  \\ 
& \left. +\displaystyle\frac{\delta \Sigma }{\delta \phi ^{*a}}\displaystyle%
\frac \delta {\delta \phi ^a}+\displaystyle\frac{\delta \Sigma }{\delta \xi
_\mu ^a}\displaystyle\frac \delta {\delta \xi ^{*a\mu }}+\displaystyle\frac{%
\delta \Sigma }{\delta \xi ^{*a\mu }}\displaystyle\frac \delta {\delta \xi
_\mu ^a}\right) \;,
\end{array}
\label{bf-lin}
\end{equation}
and

\begin{equation}
\mathcal{B}_\Sigma \mathcal{B}_\Sigma =0\;.  \label{bf-nilp}
\end{equation}
Let us turn now to the invariant vector cocycle $\left( \ref{vec-cocycle}%
\right) $, which in the present case takes the form 
\begin{equation}
\Omega _\nu ^{-1}\equiv \int d^4x\left( c^a\partial _\nu C^{*a}-A_\mu
^a\partial _\nu A^{*a\mu }+\xi _\mu ^a\partial _\nu \xi ^{*a\mu }-\phi
^a\partial _\nu \phi ^{*a}-\frac 12\;B_{\mu \tau }^a\partial _\nu B^{*a\mu
\tau }\right) \;,  \label{bf-vect-cocycle}
\end{equation}
and 
\begin{equation}
\mathcal{B}_\Sigma \Omega _\nu ^{-1}=P_\nu \Sigma =0\;.  \label{bf-vec-in}
\end{equation}
Again, the vanishing of the cohomology of the operator $\mathcal{B}_\Sigma $ 
\cite{gms,book} in the sector of the integrated local polynomials with ghost
number -1 and with a free Lorentz index implies that, as in the case of the
Chern-Simons model, $\Omega _\nu ^{-1}$ is an exact $\mathcal{B}_\Sigma $%
-cocycle 
\begin{equation}
\Omega _\nu ^{-1}=\mathcal{B}_\Sigma \Xi _\nu ^{-2}\;,  \label{bf-triv}
\end{equation}
for some local integrated polynomial $\Xi _\nu ^{-2}$ of dimension 5 and
ghost number -2. In fact, repeating the same procedure done in the previous
exemple, $\Xi _\nu ^{-2}$ is easily found to be

\begin{equation}
\Xi _\nu ^{-2}=\int d^4x\;\left( C^{*a}A_\nu ^a+\;\frac 12\varepsilon
_{\sigma \tau \mu \nu }A^{*a\sigma }B^{*a\tau \mu }-\phi ^{*a}\xi _\nu
^a-\xi ^{*a\mu }B_{\mu \nu }^a\right) \;.  \label{bf-xi-cocycle}
\end{equation}
Finally, converting the equation $\left( \ref{bf-triv}\right) $ into contact
terms by means of the expression $\left( \ref{bf-lin}\right) $ and moving
from the reduced action $\left( \ref{bf-red-action}\right) $ to the complete
one $\left( \ref{bf-comp-act}\right) $, we get the linearly broken vector
Ward identity

\begin{equation}
\mathcal{W}_\nu \mathcal{S=}\Delta _\nu ^{cl}\;,  \label{bf-susy}
\end{equation}
with 
\begin{equation}
\begin{array}{ll}
\mathcal{W}_\nu =\displaystyle\int d^4x & \left( \displaystyle\frac
12\varepsilon _{\sigma \tau \mu \nu }\left( \hat{B}^{*a\tau \mu }+\partial
^\tau \overline{\xi }^{a\mu }-\partial ^\mu \overline{\xi }^{a\tau }\right) %
\displaystyle\frac \delta {\delta A_\sigma ^a}+A_\nu ^a\displaystyle\frac
\delta {\delta c^a}-\partial _\nu \overline{c}^a\displaystyle\frac \delta
{\delta b^a}\right. \\ 
& -\displaystyle\frac 12\varepsilon _{\sigma \tau \mu \nu }\left( \hat{A}%
^{*a\sigma }+\partial ^\sigma \overline{c}^a\right) \displaystyle\frac
\delta {\delta B_{\tau \mu }^a}-B_{\mu \nu }^a\displaystyle\frac \delta
{\delta \xi _\mu ^a}-\xi _\nu ^a\displaystyle\frac \delta {\delta \phi ^a}+%
\overline{\phi }^a\displaystyle\frac \delta {\delta \overline{\xi }^{a\mu }}
\\ 
& -\partial _\nu \overline{\phi }^a\displaystyle\frac \delta {\delta \omega
^a}+\partial _\nu e^a\displaystyle\frac \delta {\delta \lambda ^a}-\left(
\omega ^a\delta _\nu ^\tau +\partial _\nu \overline{\xi }^{a\tau }\right) %
\displaystyle\frac \delta {\delta h^{a\tau }}+C^{*a}\displaystyle\frac
\delta {\delta \hat{A}^{*a\nu }} \\ 
& \left. +\phi ^{*a}\displaystyle\frac \delta {\delta \hat{\xi}^{*a\nu }}-%
\hat{\xi}^{*a\mu }\displaystyle\frac \delta {\delta \hat{B}^{*a\mu \nu
}}\right) \;,
\end{array}
\;  \label{bf-W-op}
\end{equation}
and the classical breaking $\Delta _\nu ^{cl}$ given by 
\begin{equation}
\begin{array}{ll}
\Delta _\nu ^{cl}=\displaystyle\int d^4x\; & \left( \hat{A}^{*a\mu }\partial
_\nu A_\mu ^a-C^{*a}\partial _\nu c^a-\hat{\xi}^{*a\mu }\partial _\nu \xi
_\mu ^a+\phi ^{*a}\partial _\nu \phi ^a+\displaystyle\frac 12\hat{B}^{*a\tau
\mu }\partial _\nu B_{\tau \mu }^a\right. \\ 
& \left. -\displaystyle\frac 12\varepsilon _{\sigma \tau \mu \nu }\hat{B}%
^{*a\tau \mu }\partial ^\sigma b^a+\displaystyle\frac 12\varepsilon _{\sigma
\tau \mu \nu }\hat{A}^{*a\sigma }\partial ^\tau h^{a\mu }\right) \;.
\end{array}
\label{bf-breaking}
\end{equation}
The equation $\left( \ref{bf-susy}\right) $ is recognized to be the well
known topological vector susy Ward identity of the four dimensional $BF$
systems \cite{gms}. Let us conclude by remarking that the above construction
can be easily generalized to the $BF$ systems in higher space-time
dimensions, reproducing then the results of \cite{gms,book}. In particular
it is not difficult to check that, as it happens in the case of the three
dimensional Chern-Simons model, the requirement that the breaking term $%
\left( \ref{bf-breaking}\right) $ is at most linear in the quantum fields
completely fixes the gauge parameters that one could introduce in the gauge
fixing term $\left( \ref{bf-gauge-fix}\right) $ and which would be left free
by the Slavnov-taylor identity \cite{gms}. In other words, the relative
coefficients of the lagrangian multiplier part of the gauge fixing are
uniquely determined by the vector susy Ward identity $\left( \ref{bf-susy}%
\right) $.

\subsection{The b-c ghost system}

We present here, as the last example of a topological model, the two
dimensional $b$-$c$ ghost system whose action reads

\begin{equation}
\mathcal{S}_{bc}=\displaystyle\int dzd\overline{z}\text{ }b\overline{%
\partial }c\;,  \label{bc-action}
\end{equation}
where the fields $b=b_{zz\text{ }}$ and $c=c^z$ are anticommuting and carry
respectively ghost number -1 and +1. The action $\left( \ref{bc-action}%
\right) $ is the ghost part of the quantized bosonic string action \cite
{bbrt,gsw} and, as it is well known, is invariant under the following
nonlinear nilpotent BRST transformations

\begin{equation}
\begin{array}{c}
sc=c\partial c\;, \\ 
sb=-(\partial b)c-2b\partial c\;.
\end{array}
\label{s-bc-transf}
\end{equation}
In particular, the right hand-side of the BRST transformation of the field $%
b $ is easily identified with the component $T_{zz}$ of the energy-momentum
tensor corresponding to the action $\left( \ref{bc-action}\right) $. This
property allows for the topological interpretation of the $b$-$c$ ghost
system.

Transformations $\left( \ref{s-bc-transf}\right) $ being nonlinear, one
needs to introduce two antifields $\left( b^{*}=b_{\;\overline{z}%
}^{z*},\;c^{*}=c_{zz\overline{z}}^{*}\right) $ of ghost number 0 and -2

\begin{equation}
\mathcal{S}_{ext}=\displaystyle\int dzd\overline{z}\text{ }\left(
c^{*}c\partial c+b^{*}(c\partial b-2b\partial c)\right) \;.
\label{bc-ext-act}
\end{equation}
The complete action

\begin{equation}
\mathcal{S=S}_{bc}+\mathcal{S}_{ext}\;,  \label{bc-comp-act}
\end{equation}
obeys thus the Slavnov-Taylor identity

\begin{equation}
\int dzd\overline{z}\left( \frac{\delta \mathcal{S}}{\delta b}\frac{\delta 
\mathcal{S}}{\delta b^{*}}\;+\frac{\delta \mathcal{S}}{\delta c}\frac{\delta 
\mathcal{S}}{\delta c^{*}}\;\right) =\;\frac 12\mathcal{B}_{\mathcal{S}}%
\mathcal{S}\;=\;0\;\;,  \label{bc-Slv-Tayl}
\end{equation}
$\mathcal{B}_{\mathcal{S}}$ denoting the linearized operator

\begin{equation}
\mathcal{B}_{\mathcal{S}}=\int dzd\overline{z}\left( \frac{\delta \mathcal{S}%
}{\delta b}\frac \delta {\delta b^{*}}\;+\frac{\delta \mathcal{S}}{\delta
b^{*}}\frac \delta {\delta b}+\frac{\delta \mathcal{S}}{\delta c}\frac
\delta {\delta c^{*}}\;+\frac{\delta \mathcal{S}}{\delta c^{*}}\frac \delta
{\delta c}\right) \;,  \label{bc-lin}
\end{equation}
and 
\begin{equation}
\mathcal{B}_{\mathcal{S}}\mathcal{B}_{\mathcal{S}}=0\;.  \label{bc-nilp-lin}
\end{equation}
\begin{table}[tbh]
\centering
\begin{tabular}{|c|c|c|c|c|}
\hline
& $c$ & $b$ & $c^{*}$ & $b^{*}$ \\ \hline
$dim$ & $0$ & $1$ & $1$ & $0$ \\ 
$gh-numb$ & $1$ & $-1$ & $-2$ & $0$ \\ \hline
\end{tabular}
\caption[t2]{dimension and ghost number}
\label{bc-table}
\end{table}
\\ Concerning now the vector cocycle $\left( \ref{vec-cocycle}\right) $,
here written in components, we have

\begin{equation}
\begin{array}{c}
\Omega _z^{-1}=\displaystyle\int dzd\overline{z\text{ }}\left( b\partial
b^{*}+c\partial c^{*}\right) \;, \\ 
\Omega _{\overline{z}}^{-1}=\displaystyle\int dzd\overline{z\text{ }}\left( b%
\overline{\partial }b^{*}+c\overline{\partial }c^{*}\right) \;,
\end{array}
\label{bc-vec-cocyc}
\end{equation}
and

\begin{equation}
\begin{array}{c}
\mathcal{B}_{\mathcal{S}}\Omega _z^{-1}=P_z\mathcal{S}=0\;, \\ 
\mathcal{B}_{\mathcal{S}}\Omega _{\overline{z}}^{-1}=P_{\overline{z}}%
\mathcal{S}=0\;.
\end{array}
\label{bc-cocy-inv}
\end{equation}
As done before, we look then at the solution of the BRST exact conditions

\begin{equation}
\begin{array}{c}
\Omega _z^{-1}=\mathcal{B}_{\mathcal{S}}\Xi _z^{-2}\;, \\ 
\Omega _{\overline{z}}^{-1}=\mathcal{B}_{\mathcal{S}}\Xi _{\overline{z}%
}^{-2}\;,
\end{array}
\label{bc-exact-coc}
\end{equation}
for some local integrated polynomials $\left( \Xi _z^{-2},\Xi _{\overline{z}%
}^{-2}\right) $ of ghost number -2 and dimension 1. After some almost
trivial algebraic manipulations one easily find

\begin{equation}
\begin{array}{l}
\Xi _z^{-2}=-\displaystyle\int dzd\overline{z\text{ }}c^{*}\;, \\ 
\Xi _{\overline{z}}^{-2}=-\displaystyle\int dzd\overline{z\text{ }}%
c^{*}b^{*}\;.
\end{array}
\label{bc-x-cocycles}
\end{equation}
Converting then the equations $\left( \ref{bc-exact-coc}\right) $ into
contact terms, we get the two linearly broken Ward identities

\begin{equation}
\begin{array}{l}
\displaystyle\int dzd\overline{z}\displaystyle\frac{\delta \mathcal{S}}{%
\delta c}=-\displaystyle\int dzd\overline{z\text{ }}\left( b\partial
b^{*}+c\partial c^{*}\right) \;, \\ 
\displaystyle\int dzd\overline{z}\left( b^{*}\displaystyle\frac{\delta 
\mathcal{S}}{\delta c}+c^{*}\displaystyle\frac{\delta \mathcal{S}}{\delta b}%
\right) =-\displaystyle\int dzd\overline{z\text{ }}\left( b\overline{%
\partial }b^{*}+c\overline{\partial }c^{*}\right) \;,
\end{array}
\label{bc-vec-susy}
\end{equation}
which are nothing but the topological vector susy Ward identities of the $b$-%
$c$ ghost system \cite{womss,zcurv,book}.

\section{The case of the rigid gauge invariance: the Landau Ghost Equation}

In this section we shall analyse another interesting example of a
classically linearly broken Ward identity whose contact terms can be
characterized in the same purely algebraic cohomological way of the
topological vector susy. This Ward identity is not related to a specific
gauge model, being present in all the cases in which the rigid gauge
invariance is an exact symmetry of the action. However in order to present
its derivation in a detailed way we shall consider, as explicit example, the
four dimensional pure Yang-Mills theory of the Sect.3, the generalization to
other models being straightforward.

As it is well known the rigid gauge invariance is an exact symmetry of the
Yang-Mills action $\left( \ref{red-ym-action}\right) $, expressing the
simple fact that all the fields and antifields belong to the adjoint
representation of the gauge group, \textit{i.e.}

\begin{equation}
\mathcal{R}_a^{rig}\Sigma =\int d^4x\text{ }f_{abc}\left( A_\mu ^b\frac{%
\delta \Sigma }{\delta A_\mu ^c}+A_\mu ^{*b}\frac{\delta \Sigma }{\delta
A_\mu ^{*c}}\;+c^b\frac{\delta \Sigma }{\delta c^c}+C^{*b}\frac{\delta
\Sigma }{\delta C^{*c}}\;\right) =\;\;0\;\;,  \label{ym-rig-inv}
\end{equation}
Let us consider now the following integrated local polynomial, linear in the
(shifted) antifields $\left( A_\mu ^{*},\;C^{*}\right) $, of ghost number
-1, dimension four, and possessing a group index belonging to the adjoint
representation

\begin{equation}
\Omega _a^{-1}=\int d^4x\text{ }f_{abc}\left( A^{b\mu }A_\mu
^{*c}-c^bC^{*c}\right) \;.  \label{rig-cocycle}
\end{equation}
As in the case of the vector cocycle of the eq.$\left( \ref{vec-cocycle}%
\right) $, the above cocycle turns out to be BRST invariant. In fact we have

\begin{equation}
\mathcal{B}_\Sigma \Omega _a^{-1}=\mathcal{R}_a^{rig}\Sigma =0\;,
\label{rig-cocyc-inv}
\end{equation}
due to the rigid invariance $\left( \ref{ym-rig-inv}\right) $. The operator $%
\mathcal{B}_\Sigma $ appearing in the above equation is the usual linearized
Slavnov-Taylor operator defined in eq.$\left( \ref{y-m-linear}\right) $.
However, unlike the vector cocycle $\Omega _\nu ^{-1}$, the coloured cocycle 
$\Omega _a^{-1}$ turns out to be always $\mathcal{B}_\Sigma $-exact, due to
a very well known result of the BRST cohomolgy stating that there are no
nontrivial cohomology classes with one free group index \cite{ymcoh}.
Therefore we can write

\begin{equation}
\Omega _a^{-1}=\mathcal{B}_\Sigma \Xi _a^{-2}\;,  \label{rig-exact}
\end{equation}
for some local integrated polynomial $\Xi _a^{-2}$ of ghost number -2 and
dimension four. From the Table 1 it follows that the most general form for $%
\Xi _a^{-2}$ can be written as 
\begin{equation}
\Xi _a^{-2}=\beta \int d^4x\text{ }C_a^{*}\;,  \label{xi-express}
\end{equation}
$\beta $ being an arbitrary free parameters. Using the expression of the
operator $\mathcal{B}_\Sigma $ of eq.$\left( \ref{y-m-linear}\right) $ and
of the reduced Yang-Mills action $\Sigma $ given in eq.$\left( \ref
{red-ym-action}\right) $, the condition $\left( \ref{rig-exact}\right) $
becomes

\begin{equation}
\int d^4x\text{ }f_{abc}\left( A^{b\mu }A_\mu ^{*c}-c^bC^{*c}\right) =\beta
\int d^4x\frac{\delta \Sigma }{\delta c^a}=\beta \int d^4x\;f_{abc}\left(
A^{b\mu }A_\mu ^{*c}-c^bC^{*c}\right) \;,  \label{beta-value}
\end{equation}
which gives $\beta =1$. Moving now from the reduced action $\Sigma $ to the
complete Yang-Mills action $\mathcal{S}$ $\left( \ref{red-ym-action}\right) $
and making use of the gauge condition $\left( \ref{feynm-gauge-fix}\right) $
and of the definition $\left( \ref{shifted-antif}\right) $, the equation $%
\left( \ref{beta-value}\right) $ takes the form 
\begin{equation}
\int d^4x\left( \frac{\delta \mathcal{S}}{\delta c^a}-f_{abc}\overline{c}^b%
\frac{\delta \mathcal{S}}{\delta b^c}\right) \;=\;\int d^4x\;f_{abc}\left(
A^{b\mu }\hat{A}_\mu ^{*c}-c^bC^{*c}\right) +\alpha \int d^4x\;f_{abc}b^b%
\overline{c}^c\;.\;\;\;  \label{quadr-break}
\end{equation}
It should be remarked that the left hand side of this equation, besides a
pure linear breaking, contains a term which is quadratic in the quantum
fields, \textit{i.e. }$\alpha f_{abc}b^b\overline{c}^c$. This term, being
subject to renormalization, would have been defined as an insertion,
spoiling then the usefulness of the eq.$\left( \ref{quadr-break}\right) $.
Therefore we see that the requirement that the breaking term is at most
linear in the quantum fields implies the vanishing of the gauge parameter, 
\textit{i.e. }$\alpha =0$, selecting thus the Landau gauge as the gauge
fixing condition. Finally, setting $\alpha =0$ in the gauge condition $%
\left( \ref{feynm-gauge-fix}\right) $, we obtain the linearly broken
identity 
\begin{equation}
\int d^4x\left( \frac{\delta \mathcal{S}}{\delta c^a}-f_{abc}\overline{c}^b%
\frac{\delta \mathcal{S}}{\delta b^c}\right) \;=\;\int d^4x\;f_{abc}\left(
A^{b\mu }\hat{A}_\mu ^{*c}-c^bC^{*c}\right) \;,  \label{Landau-W-id}
\end{equation}
which is recognized to be the so-called \textit{ghost equation} Ward
identity \cite{bps}, always present in the Landau gauge. Let us conclude by
recalling that the ghost equation $\left( \ref{Landau-W-id}\right) $,
although valid only in the Landau gauge, turns out to be a very powerful
tool in order to study the ultraviolet finiteness properties of a large
class of gauge invariant local field polynomials belonging to the BRST
cohomology \cite{c5}. These gauge invariant polynomials can be promoted at
the quantum level to local insertions whose anomalous dimensions are
independent from the gauge fixing parameter $\alpha $. Therefore they can be
studied without loss of generality in the Landau gauge. In particular, the
ghost equation $\left( \ref{Landau-W-id}\right) $ allows to prove the
vanishing of the anomalous dimensions of the invariant ghost monomials%
\footnote{%
We have used here the matrix notation $c=c^aT_a$, $T_a$ being the generators
of the gauge group, i.e. $[T_a,T_b]=if_{ab}^{\;\;\;c}T_c$.} $tr(c^{2n+1})$,
which are deeply related to the gauge anomalies and to the generalized
Chern-Simons terms \cite{book}. This result is of great importance in order
to prove the Adler-Bardeen nonrenormalization theorem for the gauge and the $%
U(1)$ axial anomalies \cite{c5,adb,book}. Let us mention, finally, that the
ghost equation $\left( \ref{Landau-W-id}\right) $ has been proven to be
renormalizable \cite{bps,book} and that, recently, has been extended to the $%
N=1$ supersymmetric gauge theories in superspace \cite{sl}.

\section{Conclusion}

A purely algebraic characterization of the topological vector susy and of
the Landau ghost equation Ward identities has been given. These Ward
identities, always linearly broken, are obtained by exploiting the BRST
exactness condition of antifield dependent cocycles with ghost number -1.
Applications to other kinds of linearly broken Ward identities as well as to
other topological theories and to superspace supersymmetric models are under
investigation.

\section{Acknowledgements}

The Conselho Nacional de Desenvolvimento Cient\'{\i}fico e Tecnol\'{o}gico
CNPq-Brazil is gratefully acknowledged for the financial support.

\end{document}